\documentclass[%
twocolumn,pra,
superscriptaddress, 
 amsmath,amssymb,
 aps,
]{revtex4-1}

\usepackage{graphicx}
\usepackage{array}
\usepackage{dcolumn}
\usepackage{bm}
\usepackage{listings}
\usepackage{scalefnt}
\usepackage{xcolor}
\usepackage{multirow}
\usepackage{siunitx}
\usepackage{booktabs}
\usepackage{amsthm,amsfonts}
\usepackage{adjustbox}
\usepackage{tabularx}
\usepackage{lmodern}

\usepackage{braket}

\begin{document}

\title{Microwave boson sampling}

\author{Borja Peropadre}
\email{Email: bperopadre@fas.harvard.edu}
\affiliation{Department of Chemistry and Chemical Biology, Harvard University, Cambridge, Massachusetts 02138, United States}
\author{Gian Giacomo Guerreschi}
\affiliation{Department of Chemistry and Chemical Biology, Harvard University, Cambridge, Massachusetts 02138, United States}
\author{Joonsuk Huh}
\affiliation{Mueunjae Institute for Chemistry (MIC), Department of Chemistry, Pohang University of Science and Technology (POSTECH), Pohang 790-784, Korea}
\author{Al\'an Aspuru-Guzik}
\email{Email: aspuru@chemistry.harvard.edu}
\affiliation{Department of Chemistry and Chemical Biology, Harvard University, Cambridge, Massachusetts 02138, United States}
\date{\today}

\begin{abstract}
The first post-classical computation will most probably be performed not on a universal
quantum computer, but rather on a dedicated quantum hardware. A strong candidate for achieving this
is represented
by the task of sampling from the output distribution of linear quantum optical networks.
This problem, known as boson sampling, has recently been shown to be intractable for any
classical computer, but it is naturally carried out by running the corresponding experiment.
However, 
only small scale realizations of boson sampling experiments have been demonstrated to date.
Their main limitation is related to the non-deterministic state preparation and inefficient
measurement step. Here, we propose an alternative setup to implement boson sampling that is
based on microwave photons and not on optical photons. The certified scalability of superconducting
devices indicates that this direction is promising for a large-scale implementation of
boson sampling and allows for more flexible features like arbitrary state preparation and
efficient photon-number measurements.

\end{abstract}

\maketitle


\section*{Introduction}

In the context of linear optics quantum computation, the fundamental work of Knill,
Laflamme and Milburn \cite{Knill2001} showed that a universal set of gates is
implementable when deterministic single photon sources, efficient detectors and
fast electronic feed-forward are exploited. Achieving any of these three
ingredients constitutes an impressive technological challenge by itself, but
the question is whether all these components are necessary to realize a form of
computation superior to the one of classical computers.
The answer, as explicitly put forward by Aaronson and Arkhipov
\cite{Aaronson2011}, is no. In particular, no feed-forward loops are necessary and
also the non-deterministic nature of state-of-the-art single photon sources can be
partially tolerated \cite{Motes2013,Lund2014}.

The task that Aaronson and Arkhipov proposed and showed to be, modulo a couple of
reasonable conjectures, intractable for classical computers \cite{Aaronson2011},
is the simulation of a linear optical quantum network in which the input state of
each mode corresponds to either a single photon or the vacuum state.
The problem is to sample from the photon number distribution measured at the output
of each mode. For this reason, it has been referred to as boson sampling.
Such hardness proof is remarkably important since it shows that intermediate
quantum setups can challenge the extended Church-Turing (ECT) thesis by suggesting
a physical implementation that computes more efficiently than a non-deterministic
Turing machine.
In practice, the ECT thesis is not directly refutable since it refers to an
asymptotically large scale implementation of a physical device, but the clear
indication of a scalable setup and the neat experimental demonstration of such
computation in medium-size devices would constitute a serious indication to
reconsider the ECT thesis.

The emphasis of the previous argument points to the scalability issue. In fact, the
original boson sampling setup works with optical photons that are difficult
to generate as single photons in a deterministic way and that, given the state-of-the-art, cannot be detected with almost unit
efficiency . Subsequent
proposals have suggested the use of different initial states, like two-mode squeezed
states \cite{Lund2014}, photon added/subtracted coherent states \cite{Seshadreesan2015},
or vacuum squeezed states \cite{Olson2015}. These modifications only partially solve the
bottlenecks of non-deterministic state preparation and detection efficiency making
the actual implementation of boson sampling exponentially demanding in the number of
photons \cite{Gard2014,Rohde2014}. A different approach to overcome such difficulties
is the use of alternative experimental setups. Phonons are bosonic particles
and, under corresponding Hamiltonians, behave in the same way as photons. Shen, Zhang
and Duan proposed to use trapped ions and their collective vibrations to implement boson sampling
\cite{Shen2014}. Unfortunately, the required interactions are not the natural ones
for the setup considered, so frequent and localized laser pulses are necessary to
constantly alter the dynamics with active control techniques.
This overhead limits the applicability of this construction to a small number
of vibrational modes of the trapped ions.

In this article, we propose to realize boson sampling with photons outside the optical
regime, in particular we show how microwave photons are ideal for a scalable
implementation that takes into account all three fundamental steps of the problem:
I) deterministic state preparation, II) direct implementation of the appropriate dynamics,
and III) highly efficient measurements.
In our proposal, we substitute the open-end optical waveguides with identical superconducting
resonators, one for each mode, and couple them through a superconducting ring coupler
implementing a tunable beam splitter Hamiltonian. Phase shifters are naturally implemented
by tuning the resonator frequency in an independent way with the aid of an adjacent
superconducting qubit. In this setup, state preparation is efficiently carried
out by loading the corresponding state of the qubit into each resonator using the
Jaynes-Cummings interaction in circuit quantum electrodynamics (circuit QED) \cite{blais2004,wallraff2004}.
The introduction of additional low-quality-factor (low-Q) resonators
allows the system readout through a quantum non-demolition measurement.
Note that in this proposal, the spatial degree of freedom of the waveguides is replaced
by a series of controlled steps in time evolution, as shown in Figure~\ref{fig:bs2}.
All the above operations can be performed deterministically and with high fidelity on
state-of-the-art superconducting devices with little design modifications with respect
to current setups \cite{Kelly2015}.  As demonstrated independently
in \cite{Chen14} and in \cite{Baust15, Wulschner2015}, these missing ingredients are
relatively easy to integrate in superconducting architectures. This guarantees the scalability of our proposal and
suggest superconducting platforms as a major physical candidate to the realization of
large scale boson sampling experiments.

Finally, the advantages of the proposed implementation do not only help us to address
computational complexity questions alone, even if of primary importance, they also have
a second, more practical relevance. Recently, a modified version of the original boson
sampling apparatus has been shown to be an essential component of the quantum simulation
of molecular vibronic spectra \cite{Huh2014}. 
The additional operations required to achieve such simulation are, essentially, the
application of displacement and squeezing operations.
These operations are readily carried out using superconductors by means of the manipulation
of the initial state of the photons \cite{Hofheinz08}.
Our proposal could then pave the way to the first boson sampling experiment with direct
practical implications.

\begin{figure*}[t!]
\begin{center}
\includegraphics[width=\linewidth]{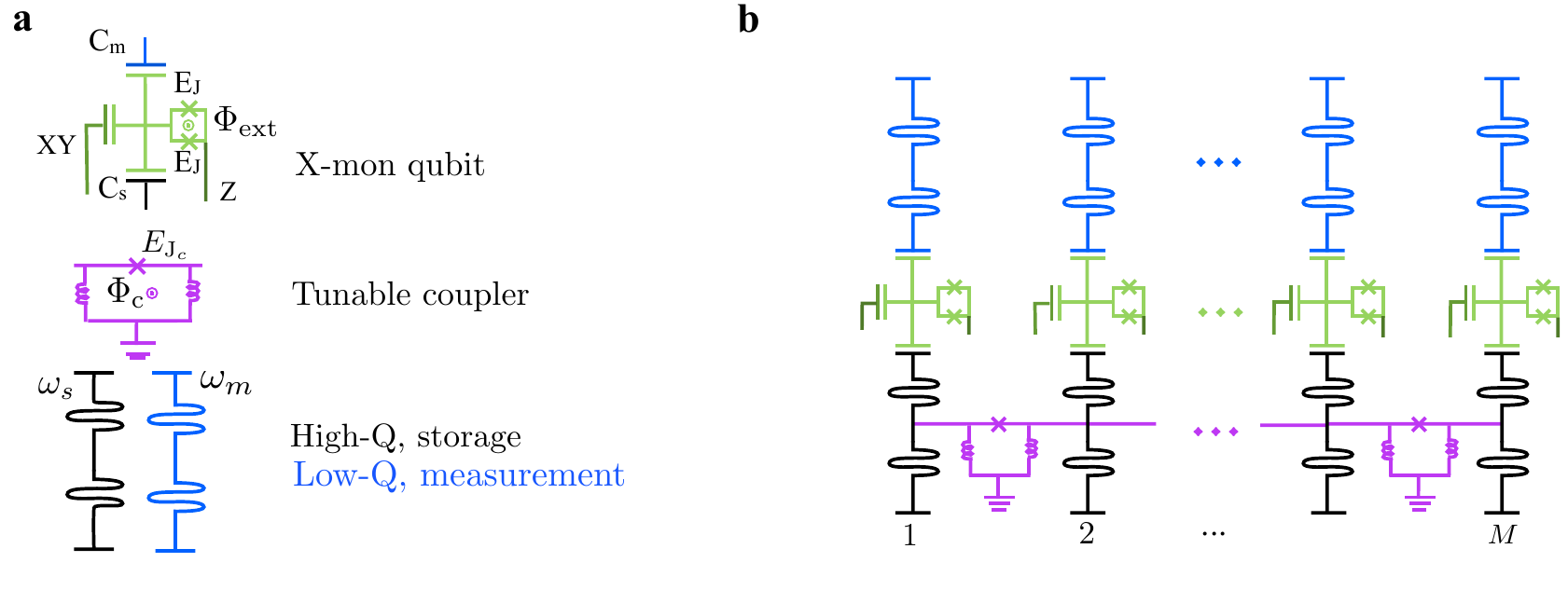}
\caption{Boson sampling implementation with microwave photons. a) Quantum electrical circuits required for the physical implementation:  (Top) a cross-shaped transmon qubit, also known as Xmon. The qubit is capacitively coupled to the storage ($C_s$) and measurement resonators ($C_m$), an $XY$ control line, and inductively coupled to a $Z$ control line that controls the qubit frequency through the external magnetic flux $\Phi_{\text{ext}}$. (Middle) A superconducting ring intersected by a Josephson junction, acting as a coupler between neighboring storage resonators. The coupling is fully tunable through the external control flux $\Phi_\text{c}$. (Bottom) High-Q storage resonator (black) and low-Q measurement resonator(blue) with characteristic frequencies $\omega_s$ and $\omega_m$. b) Experimental proposal for scalable boson sampling using superconducting circuits. }\label{fig:bs}
\end{center}
\end{figure*}

\section*{Boson Sampling Hamiltonian}
The dynamics of passive linear optical systems is determined by the sequence of beam splitter
and phase shifting elements that constitute the photonic network. Here, we show how their action
can be described in terms of the sequential application of specific Hamiltonians.

In its original formulation, boson sampling refers to the situation in which $N$ single photons
are injected in a $M$-modes photonic network characterized by the unitary matrix $U$.
Introducing the Fock number basis, \emph{i.e.} the basis composed by states $\{\ket{n_1,n_2,\cdots,n_M}\}$
having a precise number of photons $n_j$ in each mode $j=1,2,\cdots,M$, we can write the input
and output state as
\begin{eqnarray}
	\ket{\psi_{\text{in} }}&=&\ket{1_1,\cdots,1_N,0_{N+1},\cdots,0_M}  ,\\
    \ket{\psi_{\text{out}}}&=&\hat{R}_U\ket{\psi_{\text{in}}} ,
\end{eqnarray}
where the transformation $\hat{R}_U$ is defined through its action on the bosonic creation operators by
$\hat{R}_U\,a^\dagger_i\,\hat{R}_U^\dagger = \sum_j U_{ij} a^\dagger_j$.
Aaronson and Arkipov showed that sampling from the photon-number output distribution
$P(n_1,n_2,\cdots,n_M)=|\bra{n_1,n_2,\cdots,n_M}\hat{R}_U\ket{\psi_{\text{in}}}|^2$ is a computationally hard task,
provided that the number of modes $M\geq N^2$ and that the unitary $U$ is chosen
randomly according to the Haar measure \cite{Aaronson2011}.

Since any linear optical network can be constructed with phase shifters (ps) and beam splitters (bs)
alone, $\hat{R}_U$ can also be decomposed as the sequential product of the corresponding unitary
operations acting, respectively, only on one or two modes. The constructive proof that any 
$M\times M$ unitary matrix $U$ can be associated with a photonic network composed by 
$K=\mathcal{O}(M^2)$ optical elements \cite{Reck94} provides the factorization
$\hat{R}_U=\hat{U}^{(K)}\cdots\hat{U}^{(1)}$. Every operation corresponds to the
application of an appropriate Hamiltonian for the specific time $\tau_k$ according to
$\hat{U}^k=\exp{(-i\hat{H_k}\tau_k)}$. The Hamiltonians have only two possible forms
($\hbar$=1 throughout)
\begin{align}
\label{eq:bs}
\hat{H}_k^\text{bs}&= g_k a^\dagger_{i_k} a_{i_k+1}+\operatorname{H.c.} \, ,\\
\label{eq:ps}
\hat{H}_k^\text{ps}&= \phi_k a^\dagger_{j_k} a_{j_k} \, ,
\end{align}
where indexes $i_k,j_k=1,\cdots,M$ label the resonator modes involved in the $k$-th operation.
Once introduced in the operator $\hat{U}^k$, the quantities $g_k \tau_k$ and $\phi_k \tau_k$ define,
respectively, the beam splitter reflectivity and phase shift associated to the $k$-th optical element.
By applying these building-block operations sequentially, one realizes the complete boson sampling
unitary $\hat{R}_U$. This procedure offers the possibility of implementing boson sampling in any platform 
capable of generating the above Hamiltonians. In particular, superconducting circuits associate
an extraordinary level of control to the required interactions.

In the next section, we show how to implement beam splitting and phase shifting operations in circuit
QED systems. We also describe the state preparation and measurement steps that complete the scalable
implementation of boson sampling with microwave photons.

\section*{Boson sampling with superconducting circuits}
Boson sampling consists of three fundamental steps: i) initial single-photon state preparation,
ii) implementation of the random unitary $\hat{R}_U$ and iii) single-photon detection.
Here, we describe the specific circuit design to implement all the necessary operations with
microwave photons.

Our proposal consists of a series of high-quality-factor (high-Q) superconducting storage
resonators which are coupled to each other by a tunable interaction that can effectively
be switched on and off. These resonators are used for storage of the photons that will be
processed to carry out the boson sampling algorithm.
At the same time, each storage resonator is also coupled to a superconducting qubit to perform
the crucial operations required by a boson sampling device (see Figure~\ref{fig:bs}b).
While this proposal is largely qubit independent, we have chosen to illustrate
it adopting the X-mon qubit \cite{Barends13} since its cross-shaped design allows for both transverse
and longitudinal rotations without acting on the resonator themselves.
Finally, the design requires the X-mon qubit to be also coupled to a low-Q (measurement) resonator, which
will be used to perform quantum non-demolition detection of the photons stored in the storage resonator.
We now discuss how to implement all the fundamental operations in superconducting setups:

\emph{I) Initial state preparation.-} We initialize the qubits, initially far detuned in energy from
the storage resonator frequency, in the ground state $\ket{g}$. Then, we coherently drive the first
$N$ X-mon qubits through their $XY$ ports to implement a $\pi$-pulse that brings the qubits to the
excited state $\ket{e}$.
This single qubit operation can be done with extremely high fidelity, of around $99.92\%$ as recently
reported in a similar system \cite{Barends2014, Kelly2015}. By tuning the X-mon frequency through the
$Z$ qubit control line, we bring the qubits on resonance with the storage resonators for a time $t$,
activating a Jaynes-Cummings interaction of the form
\begin {equation}
\label{eq:on-res}
H_{\text{JC}}= \omega_{\text{s}} a^\dagger a+\frac{\Omega}{2}\sigma_z+g_s(\sigma^{+} a+\sigma^{-}a^\dagger)\, ,
\end{equation}
where $\Omega$ is the qubit frequency, $\omega_s$ the storage resonator frequency and $g_\text{s}$ is the coupling
constant (see Methods). Applying this interaction for a time $t=\pi/g_\text{s}$ moves the qubit excitation onto the
storage resonator $\ket{e}\otimes\ket{0}\rightarrow \ket{g}\otimes\ket{1}$, creating a single-photon Fock
state on the storage resonator.
This operation can be performed deterministically and with high efficiency, as shown in \cite{Hofheinz08}.
Interestingly enough, we are not limited to the generation of single-photon states. More complicated states,
such as higher-number Fock states \cite{Wang08} and Gaussian states \cite{Hofheinz09}, can also be prepared.
As we will discuss later on, this would allow the implementation of boson sampling with modified input states
in the form required by the quantum simulations of molecular spectroscopy \cite{Huh2014}. 

\begin{figure*}[t!]
\begin{center}
\includegraphics[width=\linewidth]{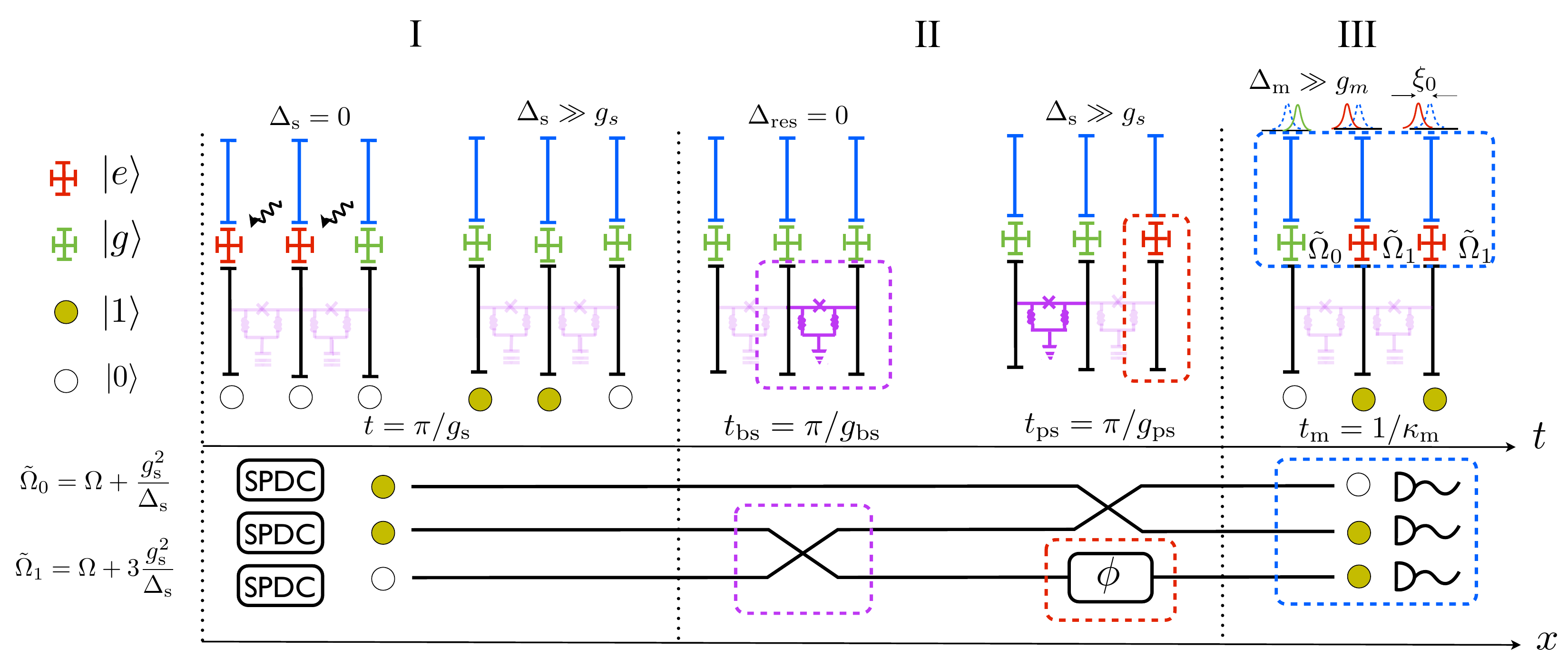}
\caption{Pictorial description of microwave boson sampling in a three-mode device, and comparison with its linear optics counterpart. In the optical network photons travel from left to right, passing through the three fundamental steps of boson sampling I) state preparation, II) unitary dynamics and III) detection. The corresponding operations in circuit QED are illustrated in the panel above, where the color code indicates which interaction is currently active. Qubits are depicted in red if in the excited state $\ket{e}$, and green if in the ground state $\ket{g}$. Purple ring coupler are disconnected when faded. The protocol is summarized in Table \ref{table1}.}\label{fig:bs2}
\end{center}
\end{figure*}

\emph{II) Unitary operation.-} In the previous section we showed that any unitary can be written as
an appropriate sequence of local Hamiltonians of the form (\ref{eq:bs}) and (\ref{eq:ps}).
Beam splitter operations can be simply carried out by bringing two transmission line resonators together.
In the confluence of their center conductors, evanescent waves couple the two resonators allowing the photons
to tunnel between them. However, their coupling is determined by the fixed geometric arrangement of the
resonators, resulting in a static coupling $g_{\text{bs}}$ that can not be switched off. In order to make
the coupling switchable, different schemes have been proposed theoretically
\cite{Peropadre13, Geller15} and implemented experimentally \cite{Chen14, Baust15, Wulschner2015}.
All these proposals are based on superconducting rings acting as tunable couplers (cf.  Fig. \ref{fig:bs}a). Switchability relies
on a controlled quantum interference between the resonator wavefunctions, that either adds them up or cancels
each other out, depending on a control parameter, namely the external magnetic flux $\Phi_{\text{c}}$
threading the superconducting ring (see Methods).
These tunable interactions have been realized both as qubit-qubit \cite{Chen14} and as resonator-resonator
couplers \cite{Baust15, Wulschner2015}, reporting on-off interaction ratios of about $10^4$. Moreover,
the switching operation is very fast and takes only a fraction of a nanosecond, that is to say a time scale
much faster than the resonator dynamics.

Phase-shifting operations can be implemented by bringing the qubit off-resonance with
the storage resonator, in the so-called dispersive regime where $\Delta_s=\Omega-\omega_s\gg g_s$. Under
this condition, the qubit induces a state-dependent pull of the resonator frequency of the form
\begin{equation}
H_{\text{dis}}=\left(\omega_s-\frac{g_s^2}{\Delta_s}\sigma_z\right)a^\dagger a+\frac{1}{2}\left(\Omega-\frac{g_s^2}{\Delta_s}\right)\sigma_z, 
\label{eq:disp}
\end{equation}
where the effective resonator frequency includes contribution from
$\phi=g_s^2/\Delta_s\times\langle\sigma_z\rangle$. As a consequence, the phase accumulated by each photon
in the resonator depends on the qubit state, being proportional to $\langle\sigma_z\rangle=\pm 1$ for the
excited and ground state, respectively. Assuming that every qubit is in the ground state $\ket{g}$,
and equally detuned with respect to its storage resonator, there is no relative frequency shift between resonators.
However, relative phases between resonators can be arbitrarily created simply by flipping the corresponding qubit
to its excited state $\ket{e}$ and introducing a frequency modification equals to $2\phi$.

Thus, applying a dispersive interaction of the form (\ref{eq:disp}) to a desired qubit-resonator pair for
times $t_{\text{ps}}\in [0,\pi/\phi]$ one can introduce arbitrary \emph{relative} phase-shifts between
any pair of adjacent storage resonators.

\begin{table*}[htb]
  \begin{adjustbox}{max width=\textwidth}
  \def\arraystretch{1.5}
  \begin{tabular}{|c||c|c|c|c|}
\hline
    \multirow{2}{*}{} &
      \multicolumn{1}{c|}{Step I: } &
      \multicolumn{2}{c|}{Step II: Unitary operator } &
      \multicolumn{1}{c|}{Step III: } \\ \cline{3-4}
     & Initial state preparation & Beam splitter & Phase sifting & Measurement protocol \\
    \hline\hline
    Physical system & qubit-storage resonator & resonator-resonator & qubit- storage resonator& qubit- measurement resonator\\
    \hline
    Hamiltonian & Jaynes-Cummings & beam-splitting  & dispersive  & dispersive  \\
    \hline
    Relevant parameters & $\Delta_\text{s}=0$, $t=\pi/g_\text{s}$ & $\Delta_{\text{res}}=0$, $t_{\text{bs}}=\pi/g_{\text{bs}}$ & $\phi=g_s^2/\Delta_{\text{s}}$, $t_{\text{ps}}=\pi/\phi$ & $\xi_0=g_m^2/\Delta_{\text{m}}$, $t_{\text{m}}=1/\kappa_m$ \\
    \hline
    Figures of merit & $g_\text{s}/2\pi\simeq 150 \text{MHz}$ & $g_{\text{bs}}/2\pi\simeq 30 \text{MHz}$ & $\phi\simeq 20\text{MHz}$, $\kappa_s/2\pi=1 \text{ KHz}$ & $\xi_0=30 \text{MHz}$, $\kappa_m/2\pi=20 \text{MHz}$  \\
    \hline
  \end{tabular}
  \end{adjustbox}
  \caption{Summary of the microwave boson sampling implementation. For each step of the protocol (columns), we display the key physical systems involved in that step, the Hamiltonian ruling the system dynamics, as well as the relevant parameters and their figures of merit. In step I, qubit and storage resonator interact of resonance via Jaynes-Cummings Hamiltonian. In step II, storage resonators are coupled on resonance via beam-splitter interaction for the purposes of beam splitting operations, while an off resonance, dispersive interaction with the qubit implements relative phase shifts. In step III, off resonance dispersive interaction, this time with the measurement resonator, is used for quantum non-demolition detection. }
  \label{table1}
\end{table*}

\emph{III) Readout.-} A very important and delicate step in any superconducting architecture is the measurement
protocol. For this reason, we provide two alternative implementations based on distinct physical mechanisms.
The first mechanism consists of mapping the storage resonator state back to the qubits, by inverting the state
preparation procedure. This mechanism is supposed to perfectly distinguish between an empty resonator and
a resonator occupied by a single microwave photon, as required in the original formulation of boson sampling
\cite{Aaronson2011}.
Bringing the qubits on resonance with the storage resonators, the interaction in eq.~(\ref{eq:on-res}) causes
Rabi oscillations that swap the boson sampling resonator state $\ket{\psi_\text{out}}$ to the qubit \cite{Hofheinz08}.
While two or more photons might have bunched together on the same resonator, thus preventing the transfer
to the qubit state due to a photon-blockade effect \cite{Ginossar10}, we can postselect this event as we would do in any
linear optics implementation.  With the aid of a second, low-Q resonator, we perform a quantum
non-demolition detection of the qubit state (see Figure \ref{fig:bs}b). Measuring the transmission of the
measurement resonator, we detect with large fidelity whether the qubits are in the ground or excited state,
and hence the photon state in the storage resonators \cite{Schuster07}.

While the measurement described above has similarity with the functioning of a ``photodetector'' (i.e. discriminate
only 0 or 1 microwave photons in the resonator), we devise a second readout mechanism that works as a
\emph{high-efficient quantum non-demolition} photon counter. The measurement mechanism is based on qubit-photon
logic gates \cite{Johnson10}. Within the dispersive regime, where the qubit is detuned by an amount $\Delta_m$,
the effective qubit frequency is lifted due to the photons in the storage resonator according to
$\tilde{\Omega}_n=\Omega+(2n+1)g_\text{s}^2/\Delta$, where $n$ is the number of photons on the storage the resonator.
Then, by sending coherent microwave signals at the different frequencies of the qubit $\tilde{\Omega}_n$,
we perform a $\pi$-rotation on the qubit, contingent on the storage resonator state $\ket{n}$: when the
driving microwave hits the qubit at its resonant frequency, we flip the qubit state $\ket{g}\rightarrow\ket{e}$,
which will, in turn, create a displacement of the measurement resonator frequency \cite{Schuster07}. By tracking
the transmission on the measurement resonator, we can determine the number of photons $n$ in the storage resonator
in at most $n$ trials. As far as boson sampling is concerned this would normally correspond to one or two attempts to measure the resonator. Each readout can be performed with efficiency of about $90\%$ \cite{Johnson10} and, since
the measurement is non-demolition, one can repeat the measurement many times to exponentially reduce the probability
of failure.
The latter readout scheme represents a remarkable feature of our microwave setup that is absent, in its deterministic
form, in linear optical setups. This results is very relevant for the realization of boson sampling
experiments that require counting more than one photon per mode, as is the case for modified boson sampling protocols
with initial Gaussian states \cite{Lund2014, Huh2014}.

To illustrate our proposal, we present in Figure~\ref{fig:bs2} a pictorial comparison of a three-mode boson sampling implementation with superconducting circuits and the original linear optical network. A summary of the whole microwave boson sampling protocol can be found in Table \ref{table1}, where we present the most relevant parameters together with their experimental benchmarks.

\section*{Generalized boson sampling with superconducting circuits}
At various points during the description of the proposed microwave setup, we observed that the superconducting
design allows not only the implementation of all passive linear elements, but also several additional operations.
This flexibility represents a necessary condition to realize many generalized versions of the boson sampling
problem \cite{Lund2014,Olson2015}.
Implementations with Gaussian states require, for example, the ability of preparing two-mode squeezed states
and to perform parity measurements. A particular role is played by the proposal in Ref.~\cite{Huh2014}, which
constitutes the first practical application of boson sampling and connects it to molecular spectroscopy.
Here, we describe how the required operations of displacement, squeezing and photon-number discrimination are achievable with microwave photons.

Consider the very same device presented to tune the resonator-resonator couplings. The specific form of the interaction
in eq.~(\ref{eq:bs}) is obtained in the rotating wave approximation starting from the more accurate form
$H_{\text{int}}= g_k(\Phi_{\text{c}})(a^\dagger_k + a_k)(a^\dagger_{k+1} + a_{k+1})$. As detailed in the methods section,
when the external magnetic flux through the coupler $\Phi_\text{c}$ oscillates at the appropriate frequency $\omega_\text{c}=\omega_k+\omega_{k+1}$, the interaction effectively
produces two-mode squeezing in the frame rotating at the coupler frequency. Simultaneously, a displacement operation can be straightforwardly introduced by simply driving the storage resonator itself.
The combined action of displacement and squeezing is interpreted as the required state preparation step of modified
boson sampling setups \cite{Lund2014, Huh2014}.
The other essential requirement is the ability of determining the parity or counting the number of photons in a resonator.
This operation has already been described in the previous section: in essence, we exploit the nearby qubit to
check a single occupation number of the storage resonator at a time, effectively implementing a quantum non-demolition photon counter. By virtue of the suggested protocol, our proposal constitutes, to the best of our knowledge, the first scalable implementation of any practical application of boson sampling.

\section*{Discussion}
To address the feasibility of our proposal, we have to understand how the requirements on the single operation
affect the overall scalability. First of all, the number of consecutive beam splitter or phase shifting operations
to be performed increases with the number of modes $M$. In general, one needs $\mathcal{O}(M^2)$ operations
\cite{Reck94}, but we observe that in our setup, like in the optical counterpart, $\mathcal{O}(M)$ operations
can be implemented simultaneously. This relaxes the requirement on the quality factor of the storage resonators,
since a resonator lifetime proportional to $M$ is sufficient. How many operations can we perform, and therefore
how many modes can we consider, before the microwave photons are lost?
Loading and measuring the resonator is performed only once per run, while a typical operation consisting of a beam splitter followed by a phase shifter requires a time $(t_\text{bs}+t_\text{ps}) \simeq 0.3 \mu s$ (see values reported in Table 1).
This time has to be compared with the storage resonator lifetime, which would probably be the limiting factor
to run a successful  experiment. High finesse coplanar waveguides resonators with quality factors above one million
have been reported \cite{Wang13a,Ohya14}, yielding cavity decay rates $\kappa\simeq 2\pi\times 1 \text{ KHz}$
corresponding to a cavity lifetime $t_\kappa= 150 \mu s$. Thus, one could implement a total number of operations
$t_\kappa/(t_\text{bs}+t_\text{ps})\simeq 500$ before the photons are lost.

Since boson sampling is believe to be hard for $N\sim\sqrt{M}$, we can successfully manipulate $\sim 20$ photons.
At the same time, the probability of correctly preparing and detecting all the $N$ single photons diminishes
exponentially in any non error-corrected architecture, and superconducting circuits are not an exception.
However, the remarkable fidelities $\mathcal{F}\simeq 99.9\%$ achieved in generating and measuring single photon Fock states \cite{Hofheinz09} demonstrate that the superconducting technology is already
mature to successfully implement boson sampling with $N\sim 20$ photons. This size is at the edge of what is
tractable on a classical supercomputer \cite{Aaronson2011,Rohde12} and, therefore, we are confident that the first
post-classical computation is within experimental reach with today's technology.

In conclusion, we propose a novel architecture to overcome the limitations exhibited by the linear optical and
ion trap implementations of boson sampling. We start from the observation that any photonic network can be decomposed
in a sequence of elementary operations generated by two kind of Hamiltonians alone. Then, we suggest to realize the
bosonic modes by identical microwave resonators that are coupled to each other with tunable strength. For each resonator,
a superconducting X-mon qubit provides the access needed to perform state preparation and measurement.
A consequence of the proposed design is that other non-linear operations, like those introduced in recent
works on generalized versions of boson sampling, are readily implementable.
In particular, squeezing operations and photon counter measurements are now available to realize the first practical
application of boson sampling in the context of molecular spectroscopy.

\section*{Methods}
\subsection*{On- and off-resonant regimes}
The X-mon qubit used in this proposal works as a split-transmon \cite{Koch07}, where a dc-SQUID acting as a tunable Josephson junction induces fast changes in the qubit frequency (see Fig. \ref{fig:bs}a, top for the circuit design, and the definition of the relevant quantities considered below). More specifically, the qubit frequency is given by
$\Omega=\sqrt{8E_J(\Phi_\text{ext})E_C},$
where $E_C=e^2/2C_J$ is the capacitive energy of the qubit, and $E_J(\Phi_\text{ext})=2E_J\cos(2\pi\Phi_\text{ext}/\Phi_0)$ is the effective Josephson energy. Through variations of the external magnetic flux $\Phi_\text{ext}$, one can change $\Omega$ to the desired frequency range. More precisely, we are interested in bringing the qubit on and off resonance with the storage resonators which accounts for frequency change of a few GHz. Such a frequency change can be done in a few nanoseconds, without altering the qubit lifetime, of the order of tens of microseconds. 

Finally, the coupling constant that characterizes the interaction between qubits and resonators, valid for both storage and measurement resonator, is given by \cite{Blais04}
\begin{equation}
g_\text{s,m}=\frac{C_{\text{s,m}}}{C_\text{s,m}+C_\text{J}} \sqrt{\frac{\omega_\text{s,m}}{cL}},
\end{equation}
where $C_{\text{s,m}}$ is the capacitive coupling of the qubit to the storage (measurement) resonator, $L$ is the resonator length, and $c$ the capacitance per unit length.
\subsection*{Beam splitting and two-mode squeezing operations}
The more generic resonator-resonator interaction can be written in the interaction picture as \cite{Peropadre13}
\begin{equation}
H_\text{int}=g(\Phi_{\text{c}})(a^\dagger_k + a_k)(a^\dagger_{k+1} + a_{k+1}),
\label{eq:int}
\end{equation}
where $a_k$ ($a^\dagger_k$) is the annihilation (creation) operator of the $k$-th resonator, satisfying canonical commutation rules $[a_k,a_{k'}^\dagger]=\delta_{kk'}$ and $g(\Phi_\text{c})$ is a flux-dependent coupling constant of the form
\begin{equation}
g(\Phi_\text{c})=g_\text{bs}\cos(2\pi\Phi_\text{c}/\Phi_0).
\end{equation}
For the purposes of implementing beam splitting operations over identical resonators ($\omega_k=\omega_{k+1}$ $\forall k$), one just needs a step-function dependence of static external fluxes, since Hamiltonian (\ref{eq:int}) follows immediately after the rotating wave approximation \cite{Peropadre13}. More precisely, for switching on a $50/50\%$ beam splitter interaction one applies an external flux $\Phi_{\text{c}}=n\Phi_0$ for a time $t=\pi/g_{\text{bs}}$, while an externally applied flux at $\Phi_{\text{c}}=(n+1/2)\Phi_0$ will switch it off.
On the other hand, the same interaction  (\ref{eq:int}) can effectively generate squeezing operations when the applied flux oscillates at the appropriate frequency. In particular, for an external flux $\Phi_\text{c}(t)=\Phi_\text{c}\cos{((\omega_k+\omega_{k+1})t)}$, and invoking the Anger-Jacobi expansion, one can go to a new rotating frame that yields the effective Hamiltonian $H_\text{int}=g_{\text{bs}}(a^\dagger_k a_{k+1}^\dagger+\operatorname{H.c})$.

\section*{Acknowledgments}
The authors would like to thank J.J. Garc\'ia-Ripoll for carefully reading the manuscript. B.P., G.G.G. and A.A.-G. acknowledge the Air Force of Scientific Research for support under award: FA9550-12-1-0046. A.A.-G. acknowledges the Army Research Office under Award: W911NF-15-1-0256 and the Defense Security Science Engineering Fellowship managed by the Office of Naval Research.
J.H. acknowledges the Mueunjae Institute for Chemistry (MIC) postdoctoral fellowship.

\section*{Author contributions}
B.P. and A.A.-G envision the project. B.P. and G.G.G. developed the protocol. B.P., G.G.G, J.H. and A.A.-G.  were involved in the research process, in discussions and in writing the manuscript.

\section*{Competing financial interests}
The authors declare no competing financial interests.




\bibliographystyle{apsrev4-1}
\bibliography{BSwSC}

\end{document}